\documentclass[aps,prd,preprint,groupedaddress,showpacs]{revtex4}
\usepackage{latexsym}
\usepackage{graphicx}
\usepackage{epsfig}
\newcommand \be {\begin{equation}}
\newcommand \bea {\begin{eqnarray}}
\newcommand \ee {\end{equation}}
\newcommand \eea {\end{eqnarray}}
\newcommand \bed {\begin{displaymath}}
\newcommand \eed {\end{displaymath}}

\newcommand{\bit}{\begin{itemize}}
\newcommand{\eit}{\end{itemize}}

\begin{document}

\title{INHOMOGENEOUS CHIRAL SYMMETRY BREAKING IN NONCOMMUTATIVE FOUR FERMION INTERACTIONS}
\author{Paolo Castorina}
\email[]{paolo.castorina@ct.infn.it}
%\thanks{}
\affiliation{Dept. of Physics, University of Catania $\;$ and  $\;$  INFN, Sezione di Catania,\\ 
Via S. Sofia 64, I-95123, Catania, Italy
}
\author{Giuseppe Riccobene}
\email[]{riccobz@yahoo.it}
%\thanks{}
\affiliation{
\centerline{Scuola Superiore di Catania, via S.Paolo 73, Catania, Italy}\\
and\\
\centerline{ Dept. of Physics, University of Catania, via S.Sofia 64, Catania, Italy}
}
\author{Dario Zappal\`a}
\email[]{dario.zappala@ct.infn.it}
%\thanks{}
\affiliation{INFN, Sezione di Catania,$\;$  and  $\;$ Dept. of Physics, University of Catania,\\
via S. Sofia 64, I-95123, Catania, Italy
}
\date{\today}
\begin{abstract}
The generalization of the Gross-Neveu  model for noncommutative 3+1 space-time has been 
analyzed. We find indications that the chiral symmetry breaking occurs 
for an inhomogeneous background as in 
the LOFF phase in condensed matter.
\end{abstract}

%\keywords{}

\maketitle

\section{ INTRODUCTION}

%%%%%%

\par The idea of noncommutative space-time coordinates in 
physics dates back to the 1940's \cite{snyder}. 
Recently, due to the discovery of Seiberg and 
Witten \cite{witten} of a map (SW map) that relates noncommutative to 
commutative gauge theories, there
has been an increasing interest in studying the impact of noncommutativity on 
fundamental as well as phenomenological issues \cite{phen}.

\par Moreover,  the idea of noncommutative space coordinates has been applied 
in condensed matter and in particular to
the theory of electrons in a magnetic field 
projected to the lowest Landau level and to  the quantized  Hall effect \cite{hall}.

\par Another interesting feature of noncommutative field theories, 
also related with condensed matter, is that noncommutativity could represent a tool to 
describe the  transition between ordered and disordered phases with inhomogeneous 
order parameters.

\par In particular,  the phase structure of   $\lambda \Phi^4$ theory has 
been recently discussed \cite{camp,gubser,chen,rivelles,noi,bieten,catterall},
and, in \cite{gubser,noi,bieten,catterall}, strong indications for a phase
transition to a non-uniform stripe phase, due to noncommutativity,
 have been given.

\par Originally the transition to an inhomogeneous phase has been considered in fermionic 
system to build a new non-uniform superconducting state
in condensed matter (LOFF phase) \cite{loff}.
The interesting result is that the inhomogeneous phase
can be more stable than the homogeneous BCS state with many 
relevant phenomenological consequences.
This phenomenon has also been reconsidered 
in the analysis of the  QCD phase structure 
and  it has been proposed that, at large density, the QCD ground state is a color 
crystalline superconductor \cite{casino}
that could be found in the core of a pulsar (for a recent review see \cite{beppe}).

\par In this paper  we investigate if a noncommutative field theoretical model
for interacting fermions shows a transition to a inhomogeneous phase where the order 
parameter, i.e., the fermionic condensate, is not  constant in space-time.

\par We consider the generalization of the cutoff Gross-Neveu  (GN) model \cite{gross}
to  3+1  noncommutative coordinates and, by using
the formalism of the effective potential for composite operators, 
introduced by Cornwall, Jackiw and Tomboulis \cite{cjt} (CJT),
in the Hartree-Fock approximation we find that, due to noncommutativity i.e.,
\begin{equation}
\label{uno}
[x_{\mu},x_{\nu}]= i \theta _{\mu \nu}\; ,
\end{equation}
there are indications for a  transition to a non-uniform chiral symmetry breaking state 
similar to the  LOFF one.

\par The  paper is organized as follows. In Section II we generalize the GN model
to the noncommutative case and briefly review the CJT formalism ;
Section  III is devoted to a preliminary analysis of the occurrence
of the transition to the non-uniform  phase;
the energy difference between the two phases is computed in Section IV and 
Section V  contains some comments and the conclusions.

\section{ NONCOMMUTATIVE GROSS NEVEU MODEL  }

 In this Section we shall  summarize the formalism of the effective action for 
composite operators
(see \cite{cjt} for details) and consider the simpler generalization of the GN model to the 
noncommutative case.

For a fermionic field and for  the composite operator, as $< \bar \psi(x)  \psi(y)>$,
the CJT effective action $\Gamma (G)$ is  given by
\begin{equation}
\label{due}
\Gamma (G) ={ {- i } {\rm Tr \, Ln} {S G^{-1}} -{i } {\rm Tr} {S ^{-1} G} + 
\Gamma_{2}(G)+{i } {\rm Tr}\, 1}\; ,
\end{equation}
where $ G(x,y)$ is the full
connected propagator of the theory, $S$ is the free massless propagator
\begin{equation}
\label{tre}
i S^{-1} (x-y)= -{\partial \hskip -0.2 cm \slash }{ \delta^4 (x-y)}\; 
\end{equation}
and $ \Gamma_{2}(G)$ is given by  all two particle irreducible vacuum graphs
in the theory with propagator set equal to $G(x,y)$.
The effective action is recovered by extremizing $\Gamma (G)$ with respect to $G$
and  the Hartree-Fock approximation  corresponds to retaining only the
lowest order contribution in coupling constant to $ \Gamma_{2}(G)$ (see \cite{cjt})).

We shall apply this formalism to evaluate the effective potential for the noncommutative
generalization of the GN model which, in the commutative case,  is defined by 
the chiral symmetric Lagrangian density:
\begin{equation}
\label{quattro}
L(x)= i \bar \psi \partial \hskip -0.2 cm \slash \psi  + g (\bar \psi \psi)^2\; .
\end{equation}

The canonical generalization of the model to the noncommutative case is obtained by 
substituting the standard product
with the star (Moyal)
product,  defined as ($i,j =1,.,4$)\cite{douglas}
\bea
&&\bar \psi_\alpha \psi_\alpha \bar \psi_\beta \psi_\beta \rightarrow
\bar \psi_\alpha *\psi_\alpha *\bar \psi_\beta * \psi_\beta  = \nonumber\\
\nonumber\\
\label{cinque}
&&\exp\left \lbrack{ i \over 2} \sum_{i<j}\theta_{\mu \nu} \partial^{\mu}_{x_i} 
\partial^{\nu}_{x_j}\right \rbrack
\Bigl ( \bar \psi_\alpha (x_1) \psi_\alpha (x_2) \bar \psi_\beta (x_3) \psi_\beta 
(x_4)\Bigr ) \left |_{x_{i}=x} \right. \; .
\eea

The effect of the star product 
on the Feynman rules of the theory is an additional momentum dependence in the 
interaction vertices for the "nonplanar" 
diagrams ( see 
\cite{douglas}),  while the "planar" diagrams have the same structure of the commutative theory.
However, in the Hartree-Fock approximation of $ \Gamma_{2}(G)$, the generalization 
in Eq. (\ref{cinque}) does not introduce
any "nonplanar" diagram due to the spin structure
of the four fermion interactions and the corresponding calculation of the effective action
is not different from the commutative GN case.

\par Analogous to the noncommutative version of the $O(N)$ scalar model 
\cite{vediamo},
we can consider a more general expression for the noncommutative
four fermion interactions which,  in the planar limit, essentially reduces to the commutative GN
model,  but maintains
genuine noncommutative contributions, i.e., nonplanar diagrams, also at lowest 
order in  $ \Gamma_{2}(G)$.

\par The simplest generalization is obtained by considering the Lagrangian density
\begin{equation}
\label{sei}
L(x)= i \bar \psi \partial \hskip -0.2 cm 
\slash \psi  + g \bar \psi_\alpha * \psi_\alpha *\bar \psi_\beta* \psi_\beta
- g \bar \psi_\alpha * \bar \psi_\beta * \psi_\alpha* \psi_\beta\; .
\end{equation}

In the standard case the addition of the second term 
is trivial since it reduces to a redefinition of the coupling $g$ and  to
add a chemical potential contribution which disappears in the infinite volume limit.
However, in the noncommutative case, 
it gives to  $ \Gamma_{2}(G)$, in the Hartree-Fock approximation, also a nonplanar
term which introduces the noncommutative effects. In momentum space $ \Gamma_{2}(G)$ 
turns out to be

\begin{equation}
\label{sette}
\Gamma_{2}(G)= g [ {\rm Tr} G(p)  {\rm Tr} G(k) (1 + e^{ik \wedge p}) -2  {\rm Tr} G(p)G(k)]\; ,
\end{equation}
where $k \wedge p = k_\mu \theta_{\mu \nu} p_\nu$ 
and the traces are over all the quantum numbers.
To obtain  the previous expression for $ \Gamma_{2}(G)$, it has been assumed that the
full fermion propagator $G(x,y)$ is a translational invariant quantity.
We shall comment on this point in the following section.

\par  The breaking of the chiral symmetry requires that the solution of the
equation which minimizes the energy,
\begin{equation}
\label{otto}
{{\delta \Gamma (G)} \over {\delta G}} = 0\; ,
\end{equation}
is such that  $\lbrace G(p),\gamma_5 \rbrace \neq 0$.

It is impossible to study the transition to the new phase with the most general
class of propagators $G$ and we shall limit ourselves to a 
Rayleigh-Ritz variational 
approach \cite{cjt}, where, however,
a meaningful  ansatz for $G$ requires at least some physical 
indications on  its asymptotic behaviors.

\par First of all, let us remember that 
in the planar approximation, i.e., $\theta \Lambda^2 \rightarrow \infty$,
where the noncommutative effects essentially disappear \cite{douglas}, the 
generalization proposed in  Eq. (\ref{sei})
gives an analogous result to the standard GN model .

In this case the translational invariant full propagator can be conveniently
parametrized as \cite{cjt}

\begin{equation}
\label{nove}
G_{pl}(x,y)=i \int 
\frac{d^4 p}{(2\pi)^4} {  (p \hskip -0.2 cm \slash  + m) \over {p^2 - m^2}}{e^{-ip(x-y)}}\; ,
\end{equation}
where $m$
is a constant which is  determined by the minimum
equation  of the effective potential

\begin{equation}
\label{dieci}
m=4g\int \frac{d^4 p}{(2\pi)^4} {m \over {p^2 + m^2}}\; .
\end{equation}

\par However, for finite $\theta \Lambda^2 $,  it easy to check that  
the ansatz in Eq. (\ref{nove}) is inconsistent with the
minimum condition. Indeed, by inserting in Eq. (\ref{due}) the expression of $\Gamma_{2}(G)$
given in Eq. (\ref{sette}), and by
using the parametrization in Eq. (\ref{nove}),  the minimum equation for the mass turns out as
(in Euclidean momenta)

\begin{equation}
\label{undici}
m=4g\int \frac{d^4 p}{(2\pi)^4} {m \over {p^2 + m^2}}  (1 +2 e^{ik \wedge p})
\end{equation}
and the solution $m =$ constant $\neq 0$ is ruled out by genuine noncommutative effects.

\par Therefore, we first improve the previous ansatz in Eq. (\ref{nove})
by introducing the following, translational invariant,
parametrization of the full propagator
\begin{equation}
\label{dodici}
G(x,y)=i \int \frac{d^4 p}{(2\pi)^4} {  (p \hskip -0.2 cm \slash  + M(p^2)) 
\over {p^2 - M(p^2)^2}}{e^{-ip(x-y)}}\; ,
\end{equation}
where the explicit dependence on the momentum has been introduced  in the parametric function $M(p^2)$.

\par Then, the minimum equation  for  $M(p^2)$ is (again in Euclidean momenta)
\begin{equation}
\label{tredici}
M(p^2)=4g \int \frac{d^4 k}{(2\pi)^4} {M(k^2) \over {k^2 +M(k^2)^2}}
 +8g \int \frac{d^4 k}{(2\pi)^4} {M(k^2) \over {k^2 +M(k^2)^2}} e^{ik \wedge p}\; .
\end{equation}

To complete the Rayleigh-Ritz variational ansatz for the propagator and to evaluate the 
effective potential, one needs to know
at least the asymptotic behaviors of the solution $M(p^2)$ for large and small (Euclidean) 
momenta.

\par In Eq. (\ref{tredici}) the $p$ dependence
is due to the second integral, since the first one is a constant for any function $M(p^2)$,
which insures the convergence in the infrared region.

\par However, the noncommutative  term $e^{ik \wedge p}$  couples
the infrared and ultraviolet asymptotic behaviors: due to the strong oscillating factor,
for small $p$ the integration region is dominated by large $k$ and {\it vice-versa}.
Then one has to proceed in a self-consistent way.
One expects that, for large $p$, the noncommutative effects are negligible and a reasonable 
behavior is
\begin{equation}
\label{quattordici}
M(p^2) \rightarrow M_\theta\; ,
\end{equation}
where $M_\theta$ is a constant. Then, by Eq. (\ref{tredici}), one obtains \cite{gubser, noi}
\begin{equation}
\label{quindici}
{M(p^2)|_{p\to 0} 
\rightarrow  8g \int \frac{d^4 p}{(2\pi)^4} {M_\theta \over {p^2 +M_\theta ^2}}
e^{ik \wedge p}}\; . 
\end{equation}

\par To simplify the calculations, the antisymmetric matrix $\theta _{\mu \nu}$ is 
assumed to be of the form
\begin{equation}
\label{sedici}
{
   \theta^{\mu\nu} = \theta \pmatrix{ 0 & 1 \cr -1 & 0} \otimes
    {\bf 1}_{d/2}
  }
\end{equation}
and \cite{gubser,noi} the integration can be easily performed and it gives

\begin{equation}
\label{diciassette}
M(p^2)|_{p \to 0} \rightarrow {2g \over {\pi^2}} M_\theta {1 \over {{\theta ^2 p^2}}}\; ,
\end{equation}
which  shows the leading behavior for small $p$,
discussed in details in \cite{gubser,noi}, due to the known IR/UV connection 
\cite{douglas}.
One can selfconsistently 
verify that, by inserting Eq. (\ref{diciassette}) in the gap equation Eq. (\ref{quindici}), the leading behavior of
$M(p^2)$ for large $p$ is a constant, as initially assumed.
Then, a good ansatz for $M(p^2)$, which reproduces 
the asymptotic behaviors of the exact solution of the gap equation,
turns out to be
\begin{equation}
\label{diciotto}
M(p^2)= M_\theta \left\lbrack 1+  {2g \over {\pi^2}} {{1 \over {{\theta ^2 p^2}}}}
\right\rbrack\; .
\end{equation}

Eq. (\ref{dodici}) and Eq.(\ref{diciotto}) represent the  Rayleigh-Ritz variational parametrization of 
the full propagator and the constant parameter
$M_\theta$ has to be determined by minimizing the energy density. 
In this translational invariant case the relation between the energy density 
$E$ and the effective action 
$\Gamma_{TI}$ is well known \cite{cjt} and one has 
\bea
&&-E=\frac{\Gamma_{TI}}{V}
=\int \frac{d^4 p}{(2\pi)^4} \left [ 2\, {\rm ln}\left (1+{{M(p^2)^2} \over {p^2}}\right ) -4 {{M(p^2)^2} 
\over {p^2+{{M(p^2)^2}}}}\right ]
+\nonumber\\ 
&&
8g \int \frac{d^4 p}{(2\pi)^4}\int \frac{d^4 k}{(2\pi)^4} {{M(p^2)M(k^2)} 
\over {[p^2+M(p^2)^2][k^2+M(k^2)^2]}}
\left (1+ 2 e^{ik \wedge p}\right )\; ,
\label{dicianove}\eea
where $V$ is the four-dimensional volume.

As in the GN model, one finds the chiral
symmetry breaking for  $g \Lambda ^2$ larger than some 
critical value $(g \Lambda ^2)_c$.
The parameter $M_\theta$ depends on the coupling constant and on 
$\theta \Lambda ^2$, and, for $\theta \Lambda^2 \rightarrow \infty$,
$M_\theta  \rightarrow m$.

\par However, the singular behavior of $M(p^2)$ for small $p$ suggests \cite{braz} a 
possible non-uniform
background and , as we shall discuss in the next section,
the translational invariant propagator used so far should  be considered as an approximation
of a more deep dynamics.

\section {INDICATIONS FOR AN INHOMOGENEOUS CHIRAL SYMMETRY BREAKING PHASE}

\par As observed in the previous section, the leading behavior for small $p$ of
$M(p^2)$ is $ \simeq  {1/(\theta ^2 p^2)}$. This signals ( despite 
of the translational invariant
approximation) that the one particle irreducible (1PI) two point function is singular 
as $p \rightarrow 0$
and this physically amounts to a long range frustration:  
$<\bar \psi(x)  \psi(x)>$ oscillates in sign
for large $x$ \cite{braz,gubser,noi}. Then the possible phase transition
should be to an ordered inhomogeneous phase, where translational invariance is broken
and the noncommutativity requires a nonuniform
order parameter and a more general ansatz for the full propagator with respect to Eq. (\ref{dodici}).

\par In the general case the order parameter is given by  ($\alpha$ is the spinorial index)
\begin{equation}
\label{venti}
<\bar \psi(x)_\alpha   \psi(x)_\alpha >=\int \frac{d^4 p}{(2\pi)^4}\int \frac{d^4 k}
{(2\pi)^4} e^{-ipx}e^{+ikx}
G_{\alpha \alpha }(p,k)
\end{equation}
and it is a constant for  the translational invariant case, i.e.,  $G(p,q) =\delta ^4 (p-k)G(p)$.
On the other hand, in the planar limit one has 
\begin{equation}
\label{ventuno}
\lim_{\theta \Lambda^2 \rightarrow \infty} G(p,k) =\delta ^4 (p-k)G_{pl}(p)\; ,
\end{equation}
where $G_{pl}(p)$ is the translational invariant solution of the planar theory in Eq. (\ref{nove}).

 \par Then, if one analyzes the problem for finite and large  $\theta \Lambda^2$,
 where the noncommutative effects start
( let us remember that in cutoff unit $p/ \Lambda,\; k/ \Lambda < 1$ ),
one can use the following approximation for  $G(p,k)$
\begin{equation}
\label{ventidue}
G(p,k) \simeq \delta ^4 (p-k)G_{TI}(p) +   F_\theta(p,k)\; ,
\end{equation}
where $ G_{TI}(p)$ is a translational invariant function which depends on $\theta \Lambda^2$
and reduces to $G_{pl}(p)$ for ${\theta \Lambda^2 \rightarrow \infty}$, and
\begin{equation}
\label{ventitre}
\lim_{\theta \Lambda^2 \rightarrow \infty} F_\theta(p,k)  = 0\; .
\end{equation}

\par Now, for $x \rightarrow \infty$, 
the dominant contribution to $<\bar \psi(x)_\alpha   \psi(x)_\alpha>$ comes from the region
$p \simeq k$ i.e.,  

\begin{equation}
\label{nuova1}
<\bar \psi(x)  \psi(x)>  \simeq \int \frac{d^4 p}{(2\pi)^4} [G_{TI}(p) +  F_\theta(p,p)]\; .
\end{equation}

In other words, a translational invariant approximation 
mimics the right behavior for large $x$ and, 
for large but finite $\theta$, there is only a small
deviation from the planar theory. 
Then, the results of the previous section give a good starting point to describe 
the fermionic condensate in these asymptotic regions, where one expects oscillating 
corrections to the constant background.
This suggests the following form of 
the full non-translational invariant propagator to the order $1/\theta^4$,
in Euclidean momenta
\bea
&&
 G_\theta(p,k) ={{(-p \hskip -0.2 cm \slash +M(p^2))\delta ^4 (p-k) }\over  {p^2 + M(p^2)^2}}+
\nonumber\\
&&
{1\over 2} \left 
(\delta ^4 (p-k-P) +\delta ^4 (p-k+P) \right ) \left( P^2 A(p,k)+P^4 B(p,k)\right )\; ,
\label{nuova2}
\eea
where all quantities are expressed in cut-off units,
the four-vector $P={{\widehat P}\over \theta}$  and $|\widehat P|=1$,
$M(p^2)$ is given by Eq. ({\ref{diciotto}) and $ A(p,k)$ 
and  $B(p,k)$ are, at this stage, generic functions.

By replacing this propagator 
in Eq. (\ref{due}), in the Hartree-Fock approximation with  interaction given in 
Eq. (\ref{sei}), it turns out that, to order $O(1/\theta^4)$, 
\begin{equation}
\label{nuova3}
\Gamma(G)=\Gamma(G)_{TI}+ P^4 (\Delta \Gamma)_{NT}\; ,
\end{equation}
where the non-translational 
invariant correction $(\Delta \Gamma)_{NT}$ depends only on the function 
$ A(p,k)$, which we choose as 
\begin{equation}
\label{nuova4}
A(p,k)={{-k \hskip -0.2 cm \slash +M(k^2) }\over  {k^2 + M(k^2)^2}}\; ,
\end{equation}
to preserve the spin  structure of the translational invariant 
propagator \cite{nota1}.

After a straightforward calculation one obtains the following form of the correction
\begin{equation}
\label{nuova5}
\frac{(\Delta \Gamma)_{NT}}{V}=
4 g P^4 \int \frac{d^4 p}{(2\pi)^4}\int \frac{d^4 k}{(2\pi)^4} {{M(p^2)M(k^2)} 
\over {[p^2+M(p^2)^2][(k^2+M(k^2)^2]}}
e^{-i(p+k) \wedge P}( 1+2 e^{2 i p \wedge k})\; .
\end{equation}
One should note that the remarkable factorization of the volume factor $V$, which follows from 
the  ansatz in Eq. ({\ref{nuova2}), despite of its non-translational invariance, implies that 
the right hand side of Eq. (\ref{nuova5}) is an energy density.

For $P\to 0$, 
due to the 
large oscillating factors in the integrands, the behavior of $(\Delta \Gamma)_{NT}$
is dominated by the integration regions of large $p$ and $k$ and it turns out that 
\begin{equation}
\label{nuova6}
\frac{(\Delta \Gamma)_{NT}}{V}\simeq \frac{g}{\theta^4}\; .
\end{equation}

This result gives the indication of a transition to a nonuniform background 
related to the nontranslational invariant 
ansatz in Eq. (\ref{nuova2}). However, the CJT effective action has a clear physical interpretation
as the energy density of the system, $E$,  only  for spacetime  
translational invariant propagators (see Eq. (\ref{dicianove})). For
static, but  not space translational invariant systems,  $-{\Gamma(G)|_{static}}= {\tau} {\cal E}_T$,
where $\tau$ is the time interval and $ {\cal E}_T$ is the total energy  \cite{cjt}.
Therefore, the study of a possible phase transition to an inhomogeneous state, due to 
noncommutative effects, should, more correctly, be performed by taking 
$\theta_{0i}=0$, $\theta_{ij}\neq 0$ with 
$i,j=1,2,3$, and by using the time independent formalism.
This is the subject of the next Section.

\section {STATIC FORMALISM}

The static formalism has been developed in \cite{cjt} only for the scalar fields and, in
Appendix A, we extend it to the case of fermionic  fields.
The static propagator can be written as 
\begin{equation}
G(\vec{x},\vec{y})=G(0,\vec{x},\vec{y})=
i\int\frac{\,d \omega}{2\pi}\left[\frac{\omega \gamma_0-\vec{\gamma}\cdot\vec{f}+m}{\omega^2-F^2}\right](\vec{x},\vec{y})\; ,
\end{equation}
where 
$F(\vec{x},\vec{y})=(f^2+m^2)^{1/2}(\vec{x},\vec{y})$ and the two functions $f(\vec{x},\vec{y})$
and $m(\vec{x},\vec{y})$ describe the general time translational invariant solution of the gap equation (see Appendix A).

The energy of the system ${\cal E}_T$ is
\begin{eqnarray}
&&{\cal E}_T=
-4m\int\,d^3{x}\, \left [ {\rm Tr_{(spin)}}  G(\vec{x},\vec{x}) \right ]^{-1}-\nonumber\\
&&\int\,d^3{x}\,
{\rm Tr_{(spin)}}  \left[ \int\frac{\,d\omega}{2\pi}\gamma_0 \, \omega  {G}(\omega,\vec{x},\vec{x})-
\left.i  \vec{\gamma}\cdot\vec{\nabla}G(\vec{x},\vec{y})\right|_{\vec{x}=\vec{y}}  \right ] -
\nonumber\\
&&\Gamma_2(G)|_{static}\; ,
\label{energy}
\end{eqnarray}
where $\Gamma_2(G)|_{static}$ corresponds to the last term in Eq. (\ref{due}), evaluated in the static limit.

By following the same steps of the four dimensional calculation of Sections II and III, 
we initially  consider,  for the static propagator in the commutative case, the ansatz
\begin{equation}
 {G}(\vec{p})=\frac{\gamma^0p^0-\vec{\gamma}\cdot\vec{p}+m}{2p^0}
\end{equation}
with constant $m$ and $p^0=\sqrt{ {\vec p}^2 +m^2} $,
which gives the gap equation of the  GN model in static limit  
\begin{equation}
m=3g\int\frac{\,d^3p}{(2\pi)^3}\frac{m}{\sqrt{\vec{p}^2+m^2}}\; .
\end{equation}

If one considers the previous ansatz for $G$, in the 
noncommutative model in  Eq. (\ref{sei}), the gap equation turns out to be
\begin{equation}
m=2g\int\frac{\,d^3p}{(2\pi)^3}\frac{m}{ \sqrt{\vec{p}^2+m^2}   } (1+2e^{i\vec{p}\wedge\vec{k}})\; ,
\end{equation}
which rules out a solution with constant $m$ and requires a more general ansatz,
where $m$ has a parametric dependence  on $\vec p$, i.e.,
\begin{equation}
 {G}(\vec{p})=\frac{\gamma^0p^0-\vec{\gamma}\cdot\vec{p}+m(\vec p)}{2\sqrt{\vec{p}^2+m(\vec p)^2}}\; .
\end{equation}

The gap equation is now 
\begin{equation}
m(\vec k)=2g\int\frac{\,d^3p}{(2\pi)^3}\frac{m(\vec p)} { \sqrt{\vec{p}^2+m(\vec p)^2}}(1+2e^{i\vec{p}\wedge\vec{k}})\; ,
\end{equation}
with a selfconsistent asymptotic  solution
\begin{equation}
m(\vec k) \to m_0 \; ,  \;\;\;   \;\;\; k\to \infty\; ,
\end{equation}
\begin{equation}
m(\vec k) \to  \frac{g m_0}{ \pi^2 |\vec k\times \vec\theta |^2} \; , \;\;\;   \;\;\;   k\to 0\; ,
\end{equation}
where $m_0$ is a constant, 
the position $\theta^{ij}=\epsilon^{ijk}\theta_k$ defines the vector $\vec \theta$ and $\times$ indicates the standard 
vector product.
With  the ansatz 
\begin{equation}
m(\vec k) = m_0  \left [ 1+ \frac{g}{ \pi^2}\frac{1}{ |\vec k\times \vec\theta |^2}\right ]
\end{equation}
and  by following the same arguments given in Section III, it is straightforward to show that
the energy of the system, evaluated by 
the nontranslational invariant ansatz for the static propagator 
\bea
\label{aaa1}
&&   G_\theta(\vec p,\vec k) =
 \frac{\gamma^0p^0-\vec{\gamma}\cdot\vec{p}+m(\vec p)}{2\sqrt{\vec{p}^2+m(\vec p)^2}}\delta ^3 (\vec p-\vec k)+
\nonumber\\
&&
{1\over 2} \left 
(\delta ^3 (\vec p-\vec k- \vec P) +\delta ^3 (\vec p-\vec k+\vec P) \right ) \left( \vec P^2 A'(\vec p,\vec k)+\vec P^4 B'(\vec p,
\vec k)\right )\; ,
\label{aaa2}
\eea
where $A'(\vec p,\vec k)$ is, analogous to the four dimensional case, 
\begin{equation}
A'(\vec p,\vec k)= \frac{\gamma^0k^0-\vec{\gamma}\cdot\vec{k}+m(\vec k)}{2\sqrt{\vec{k}^2+m(\vec k)^2}}\; ,
\end{equation}
turns out to be lower than the noncommutative translational invariant case by terms of order $g/\theta^4$.

Therefore, the previous calculation gives a clear  indication 
that the noncommutative effects are responsible 
for the occurrence of the chiral symmetry 
breaking in an inhomogeneous phase, since
the latter  has always lower energy than the (translational invariant) homogeneous one.

The qualitative agreement 
between the static calculation and the 
approach in Section III  is expected  due to the following points: 
i) we are considering only a slowly varying 
background with fluctuations amplitude suppressed by powers of $1/\theta$;
ii) the non-trivial factorization of the volume $V$ in Eq. (\ref{nuova5})
makes possible a physically meaningful 
evaluation of the non-translational invariant correction
to the energy density.

In both cases (static and non-static), the
energy density  difference between these two phases 
is of order $O(g/\theta^4)$, while the difference with respect to 
the planar theory is much larger.  
Then, for convenience, in  Figure 1 we plot $-\Gamma(G)/V$, in the non-static calculation,
for the planar  theory and for the non-translational invariant case,
and in Figure 2 it is shown that  the noncommutative effects decrease  the critical coupling
constant with respect to the planar GN model.

According to point i), one can easily evaluate the $x$ dependence of the vacuum 
condensate to order $O(1/\theta^2)$, which turns out to be 
\begin{equation}
\label{nuova7}
<\bar \psi(x)  \psi(x) > =\left  ( 1+c \, P^2 {\rm cos} \left ( Px\right )\right ) <\bar \psi  \psi >_{TI}\; ,
\end{equation}
where $c$ is a constant and $ <\bar \psi  \psi >_{TI}$ is the constant 
order parameter evaluated in the translational invariant case.

\section {COMMENTS AND CONCLUSIONS}

Our computation of the CJT effective action 
shows that the chiral symmetry breaking occurs for an inhomogeneous phase,
due to the  noncommutative nature of the four fermion interactions considered 
in Eq. (\ref{sei}).
The energy difference between the inhomogeneous and the homogeneous 
phases, which both include the noncommutative corrections, 
is of order $O(g/\theta^4)$.
The order parameter has an oscillating $x$ dependence of order $O(g/\theta^2)$,
superposed to the constant background of the translational invariant phase. 
 
These results are essentially based on the non-translational invariant ansatz 
for the full propagator in Eqs. (\ref{nuova2}) and (\ref{nuova4}). Let us notice that,
in the commutative  1+1 dimensional GN model,
non-translational invariant  effects have been introduced in \cite{unopiu}  and 
a transition  to the inhomogeneous crystal 
phase at non-zero chemical potential has been obtained.

In \cite{gubser,noi} for the noncommutative scalar case,
it has been observed that 
the boson condensation does not occur in the 
mode $k=0$ but there is a total depletion to $k=Q$
where $<\phi (x)>\propto {\rm cos} (Qx)$. Analogously, in the fermionic case, 
our ansatz in Eq. (\ref{nuova2}) corresponds to Cooper pairs with 
a non-zero total momentum, as it happens in the LOFF phase in condensed matter.
The latter point provides an indication that the noncommutative cutoff field theory 
could be applied to describe the  features of the transition to inhomogeneous
phases.

\vspace{2 cm}

\appendix*
\section{A}

In this Appendix we derive the energy for time, but not space  translational invariant fermionic systems with Lagrangian 
given in Eq. (\ref{sei}), in terms of the static propagator $G(\vec x,\vec y)$ and we closely follow the procedure outlined in \cite{cjt} 
for the scalar theory. The total energy is related  to the effective action, computed in the static limit 
\begin{equation}
{\cal E}_T(G(\vec x,\vec y))\tau=-\Gamma(G(x,y))|_{static}\; .
\end{equation} 

The static limit of the effective action is obtained  by taking  the time translational invariant  propagator $G(x,y)=G(x_0-y_0,\vec x,\vec y)$
at equal time $x_0=y_0$ and by re-expressing $\Gamma$ in terms of the static propagator, defined by  the full propagator 
as $G(\vec x,\vec y)=G(0,\vec x,\vec y)$.
To obtain the form of  $G(\vec x,\vec y)$, we recall that the functional derivative of $\Gamma$ with respect to the propagator is related,
as explained in detail in \cite{cjt}, to the bilocal  source $K(\vec x,\vec y)$
\begin{equation}
 \frac{\delta \Gamma(\phi,G)} {\delta G(x_0-y_0;\vec{x},\vec{y})}=
-\frac{1}{2} \delta (x_0-y_0)K (\vec{x},\vec{y})\; ,
\end{equation}
and therefore, from the explicit derivation of the effective action, one gets
\begin{equation}
 G^{-1}(x,y)=
-i\delta(x_0-y_0)K(\vec{x},\vec{y})+S^{-1}\delta^4(x-y)+
\frac{\delta\Gamma_2}{\delta G}\; ,
\end{equation} 
which shows that the most general form of  $G^{-1}(x,y)$ is
\begin{equation} 
\label{fgg}
G^{-1}(x_0-y_0,\vec x,\vec y)=
\delta'(x_0-y_0)\gamma_0\delta(\vec{x}-\vec{y})+i\delta(x_0-y_0)\vec{\gamma}\cdot\vec{f}(\vec x,\vec y)
+im(\vec{x},\vec{y})\; ,
\end{equation}
where $\vec{f}(\vec x,\vec y)$ and $m(\vec{x},\vec{y})$ are generic functions of the spatial coordinates
and all the dependence of $G^{-1}(x,y)$ on the temporal coordinates   is contained in the delta function 
$\delta(x_0-y_0)$ and its derivative .
From Eq. (\ref{fgg}) one gets  the Fourier transform of  $G^{-1}(x_0-y_0,\vec x,\vec y)$ with respect to the variable 
$x_0-y_0$,  
\begin{equation} \label{tfing}
 {G}^{-1}(\omega,\vec{x},\vec{y}) =
\int_{-\infty}^{+\infty}\,dx_0
e^{i\omega x_0}G^{-1}(x_0,\vec{x},\vec{y}) =
-i\omega \gamma_0\delta(\vec{x}-\vec{y})+i\vec{\gamma}\cdot\vec{f}(\vec{x},\vec{y})+im(\vec{x},\vec{y})
\end{equation}
which can be functionally inverted  
\cite{nota2} :

\begin{equation}
 {G}(\omega,\vec{x},\vec{y})=i\left(\frac{\omega\gamma_0-\vec{\gamma}\cdot\vec{f}+m}{\omega^2-F^2}\right)(\vec{x},\vec{y})\; ,
 \end{equation}
where $F(\vec{x},\vec{y})=(f^2+m^2)^{1/2}(\vec{x},\vec{y})$.

Finally, the static propagator is obtained by integration
\bea
&&
G(\vec{x},\vec{y})=G(0,\vec{x},\vec{y})=
\nonumber\\
&&i\int\frac{\,d\omega}{2\pi}\left[\frac{\omega \gamma_0-\vec{\gamma}\cdot\vec{f}+m}
{\omega^2-F^2}\right](\vec{x},\vec{y})=
\frac{(\gamma_0
F-\vec{\gamma}\cdot\vec{f}+m)}{2F}(\vec{x},\vec{y})\; .
\label{stpr}
\eea
Incidentally, we note that the trace over the spin indices gives 
\begin{equation}\label{trstpr}
{\rm Tr_{(spin)}} G (\vec{x},\vec{y})=\left ( \frac{2 m}{F}\right ) (\vec{x},\vec{y})\; .
\end{equation}

We are now able to evaluate the effective action in the static limit and, for simplicity, 
we shall restrict the following calculation 
to a constant mass  $m(\vec{x},\vec{y})=m\delta(\vec{x}-\vec{y})$.
We start considering the first term in the general expression 
of $\Gamma$ in Eq. (\ref{due}), namely  $-i {\rm Tr\, Ln} G^{-1}$, 
where the trace refers both to spacetime and spin indices
(we neglect in Eq. (\ref{due}) the logarithm of the free propagator $S$ which gives a constant
contribution to the effective action). With the help of Eq. (\ref{tfing})
one gets \cite{cjt}
\bea
&&-i \int\,dx_0 \int\,d^3x \int_{-\infty}^{+\infty}\frac{\,d\omega}{2\pi} {\rm Tr_{(spin)}\, Ln }
( {G}^{-1}(\omega,\vec{x},\vec{y}))=
\nonumber\\
&&
-2i\int\,dx_0  \int\,d^3x  \int_{-\infty}^{+\infty}\frac{\,d\omega}{2\pi}
{\rm Ln} (\omega^2-F^2)=\nonumber\\
&& 
2\int\,dx_0\int\,d^3x \, F(\vec{x},\vec{x})=4m\int\,dx_0\int\,d^3x [ {\rm Tr_{(spin)}} G (\vec{x},\vec{x}) ]^{-1}\; ,
\label{con1}
\eea
where we have used Eq. (\ref{trstpr}) to replace $F$ in the last step.

The second term to compute  in  Eq. (\ref{due}) is  $-i {\rm Tr} S^{-1}G$
which yields, after integrating by parts,
\bea
&&
-i\, {\rm Tr_{(spin)}}\int\,d^4x\,d^4y[\slash\hspace{-2mm}\partial\delta^4(x-y)]G(x_0-y_0;\vec{x},\vec{y})=
\nonumber\\
&&i\,{\rm Tr_{(spin)}} \int\,d^4x\left[\left.\int\frac{\,d\omega}{2\pi}\gamma_0(-i\omega) {G}(\omega,\vec{x},\vec{x})-
\vec{\gamma}\vec{\nabla}G(0,\vec{x},\vec{y})
\right|_{\vec{x}=\vec{y}}\right]=
\nonumber\\
&&-4m\int\,dx_0\int\,d^3{x}\,
 [ {\rm Tr_{(spin)}} G (\vec{x},\vec{x}) ]^{-1}
-\left. i {\rm Tr_{(spin)}} 
\int\,dx_0\int\,d^3{x} \, \vec{\gamma}\cdot\vec{\nabla}G(\vec{x},\vec{y})\right|_{\vec{x}=\vec{y}}\; .
\label{con2}\eea

Finally the term corresponding to $\Gamma_2$ in the Hartree-Fock approximation are straightforwardly computed by replacing
$G(x_0-y_0,\vec x, \vec y)|_{x_0=y_0}$ with $G(\vec x, \vec y)$. By collecting the various contributions to the effective action,
namely Eqs. (\ref{con1}) and (\ref{con2}) plus  $\Gamma_2$ in the Hartree-Fock approximation, we get the expression of the energy 
shown in Eq. (\ref{energy}).

\begin{acknowledgments}
We thank Roman Jackiw and So Young Pi for many fruitful suggestions. The authors acknowledge
the MIT Center for Theoretical Physics for kind hospitality.
P.C. has been partially supported by the INFN Bruno Rossi exchange program.  
\end{acknowledgments}

\vfill
\eject

\begin{figure}
\epsfig{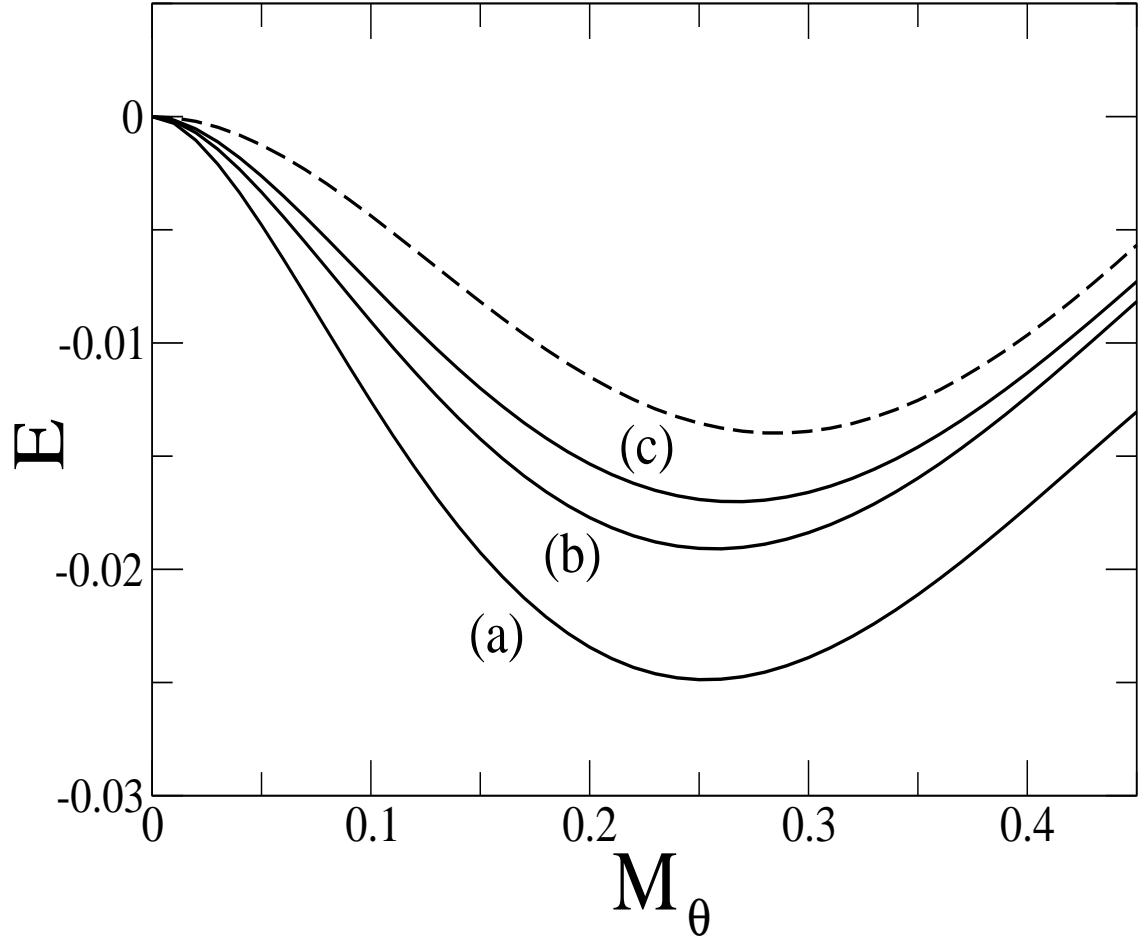}
\caption{The energy density $E$ at $(\Lambda^2 g/2\pi^2)=2.5$ 
for the planar theory (dashed line) and for the noncommutative non-translational
invariant case for $\theta\Lambda^2=12$ (a), $15$ (b), $18$ (c).
}
\end{figure}

\vfill
\eject

\begin{figure}
\epsfig{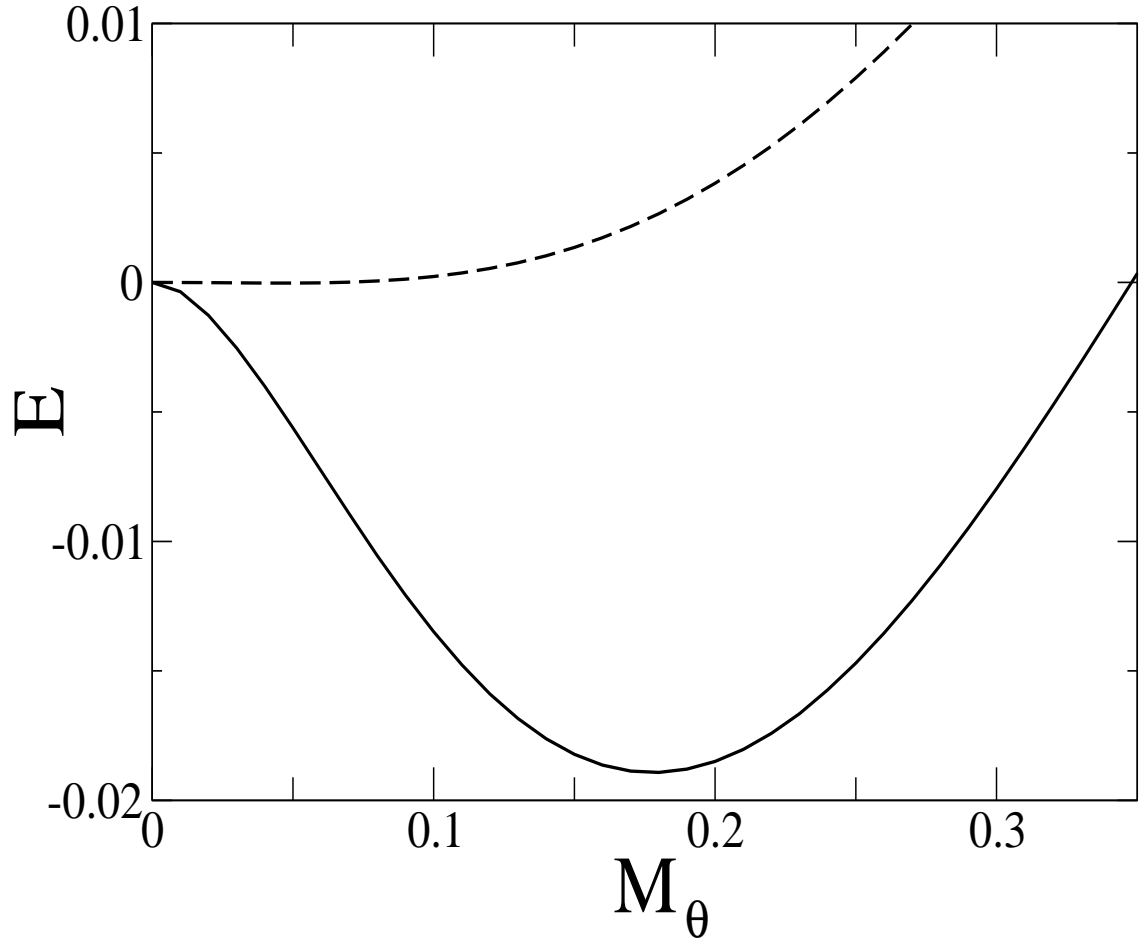}
\caption{The energy density $E$ at $(\Lambda^2 g/2\pi^2)=2 $
for the planar theory (dashed line) and for the noncommutative non-translational
invariant case for $\theta\Lambda^2=8$ (solid line).
}
\end{figure}


\begin{thebibliography}{18}
\bibitem{nota1}
To order $P^4$, the expression in Eq. (\ref{nuova4}) is equivalent to considering  the symmetric form 
$$A(p,k)=\frac{1}{2}\left ( \;{{-k \hskip -0.2 cm \slash +M(k^2) }\over  {k^2 + M(k^2)^2}} +
{{-p \hskip -0.2 cm \slash +M(p^2) }\over  {p^2 + M(p^2)^2}} \right ) 
$$
\bibitem{nota2}
This statement is not valid in general and one has to verify in each single case the commutators in functional sense 
among $\vec{f}(\vec{x},\vec{y})$ and $m(\vec{x},\vec{y})$. With the ansatz in Eq. (\ref{aaa1}),
the previous commutators vanish to  order $P^4$.
\end{thebibliography}
\end{document}